\documentclass[conference]{IEEEtran}
\usepackage{xspace}
\usepackage{xcolor}
\usepackage{amsmath, amsfonts, amsthm} 
\usepackage{graphicx}
\usepackage{float} 
\usepackage{hyperref}
\usepackage{wrapfig}
\usepackage{mathrsfs}
\usepackage{multicol}
\usepackage{balance}
\usepackage{mathtools}
\usepackage{tikz}

\usepackage{tabularx} 
\usepackage{adjustbox} 
\usepackage{multirow} 

\usepackage{complexity}
\usepackage{relsize}
\usepackage{graphicx}
\usepackage{epsfig}
\usepackage{flushend}
\usepackage{epstopdf}
\usepackage{longtable}
\usepackage{stmaryrd}
\usepackage{pstricks}
\usepackage{pst-tree}
\usepackage{epstopdf}
\usepackage{lipsum}
\usepackage{color}
\usepackage{framed}
\usepackage[normalem]{ulem}
\usepackage{float}
\usepackage{caption}
\usepackage{fancyvrb}
\usepackage{bm}

\usepackage{algorithmic}
\usepackage{algorithm}

\newcommand{\eat}[1]{}

\newcommand{\ignore}[1]{}

\newcommand{\share}[1]{\llbracket #1 \rrbracket}

\newcommand{\pprotocol}[5]{
{
\begin{center}
 \advance\leftskip-2cm
 \advance\rightskip-2cm
\setlength{\protowidth}{\textwidth}
\addtolength{\protowidth}{\intextsep}
\fbox{
        \hbox{\quad
        \begin{minipage*}{\protowidth}
        #5
        \end{minipage*}
        \quad}
        }
        \caption{\label{#3} #2}
\end{center}

} }

\renewcommand{\paragraph}[1]{\vskip 0.08in\noindent {\bf #1.}}

\usepackage{subfig}

\IEEEoverridecommandlockouts
\usepackage{cite}
\usepackage{amsmath,amssymb,amsfonts}
\usepackage{algorithmic}
\usepackage{graphicx}
\usepackage{textcomp}
\usepackage{xcolor}
\def\BibTeX{{\rm B\kern-.05em{\sc i\kern-.025em b}\kern-.08em
    T\kern-.1667em\lower.7ex\hbox{E}\kern-.125emX}}

\usepackage{lipsum}

\begin{document}

\title{PASNet: \underline{P}olynomial \underline{A}rchitecture \underline{S}earch Framework for Two-party Computation-based Secure Neural \underline{Net}work Deployment\\
}


\author{\IEEEauthorblockN{ \textsuperscript{$\star$}Hongwu Peng\textsuperscript{[1]}, \textsuperscript{$\star$}Shanglin Zhou\textsuperscript{[1]}, \textsuperscript{$\star$}Yukui Luo\textsuperscript{[2]}, Nuo Xu\textsuperscript{[3]}, Shijin Duan\textsuperscript{[2]}, Ran Ran\textsuperscript{[3]}, Jiahui Zhao\textsuperscript{[1]}, \\ Chenghong Wang\textsuperscript{[4]}, Tong Geng\textsuperscript{[5]}, Wujie Wen \textsuperscript{[3]}, Xiaolin Xu\textsuperscript{[2]}, and Caiwen Ding\textsuperscript{[1]}}
\IEEEauthorblockA{
\textsuperscript{$\star$}These authors contributed equally. \\
\textsuperscript{[1]}University of Connecticut, USA. 
\textsuperscript{[2]}Northeastern University, USA. 
\textsuperscript{[3]}Lehigh University, USA. \\ \textsuperscript{[4]}Duke University, USA. 
\textsuperscript{[5]}University of Rochester, USA. \\
\textsuperscript{[1]}\{hongwu.peng, shanglin.zhou, jiahui.zhao, caiwen.ding\}@uconn.edu,   \textsuperscript{[2]}\{luo.yuk, duan.s, x.xu\}@northeastern.edu,  \\ \textsuperscript{[3]}\{nux219, rar418, wuw219\}@lehigh.edu,  \textsuperscript{[4]}cw374@duke.edu, \textsuperscript{[5]}tgeng@ur.rochester.edu
}
}

\maketitle
\begin{abstract}
Two-party computation (2PC) is promising to enable privacy-preserving deep learning (DL). However, the 2PC-based privacy-preserving DL implementation comes with high comparison protocol overhead from the non-linear operators. This work presents PASNet, a novel systematic framework that enables low latency, high energy efficiency \& accuracy, and security-guaranteed 2PC-DL by integrating the hardware latency of the cryptographic building block into the neural architecture search loss function. We develop a cryptographic hardware scheduler and the corresponding performance model for Field Programmable Gate Arrays (FPGA) as a case study. The experimental results demonstrate that our light-weighted model PASNet-A and heavily-weighted model PASNet-B achieve 63 ms and 228 ms latency on private inference on ImageNet, which are 147 and 40 times faster than the SOTA CryptGPU system, and achieve 70.54\% \& 78.79\% accuracy and more than 1000 times higher energy efficiency. The pretrained PASNet models and test code can be found on Github\footnote{\url{https://github.com/HarveyP123/PASNet-DAC2023}}.

\begin{IEEEkeywords}
Privacy-Preserving in Machine Learning, Multi Party Computation, Neural Architecture Search, Polynomial Activation Function, Software/Hardware Co-design, FPGA
\end{IEEEkeywords}

\end{abstract}


\section{\textbf{Introduction}}

Machine-Learning-As-A-Service (MLaaS) has been an emerging solution nowadays, to provide accelerated inference for diverse applications. 
However, most MLaaS require clients to reveal the raw input to the service provider~\cite{kumar2020cryptflow} for evaluation, which may leak the privacy of users. 
Privacy-preserving deep learning (PPDL) and private inference (PI) have emerged to protect sensitive data in deep learning (DL).
The current popular techniques include multi-party computation (MPC) \cite{knott2021crypten} and homomorphic encryption (HE) \cite{kim2022secure}. 
HE 
 is mainly used to protect 
small to medium-scale DNN models without involving costly bootstrapping and large communication overhead. 
MPC protocols such as secret-sharing~\cite{knott2021crypten} and Yao's Garbled Circuits (GC)~\cite{bellare2012adaptively}
can support large-scale networks by evaluating operator blocks.   
This work mainly focuses on secure two-party computation (2PC), which represents the minimized system for multi-party computing (MPC) and is easy to extend~\cite{demmler2015aby}.

\begin{figure}[t]
    \centering
      \includegraphics[width=0.99\linewidth]{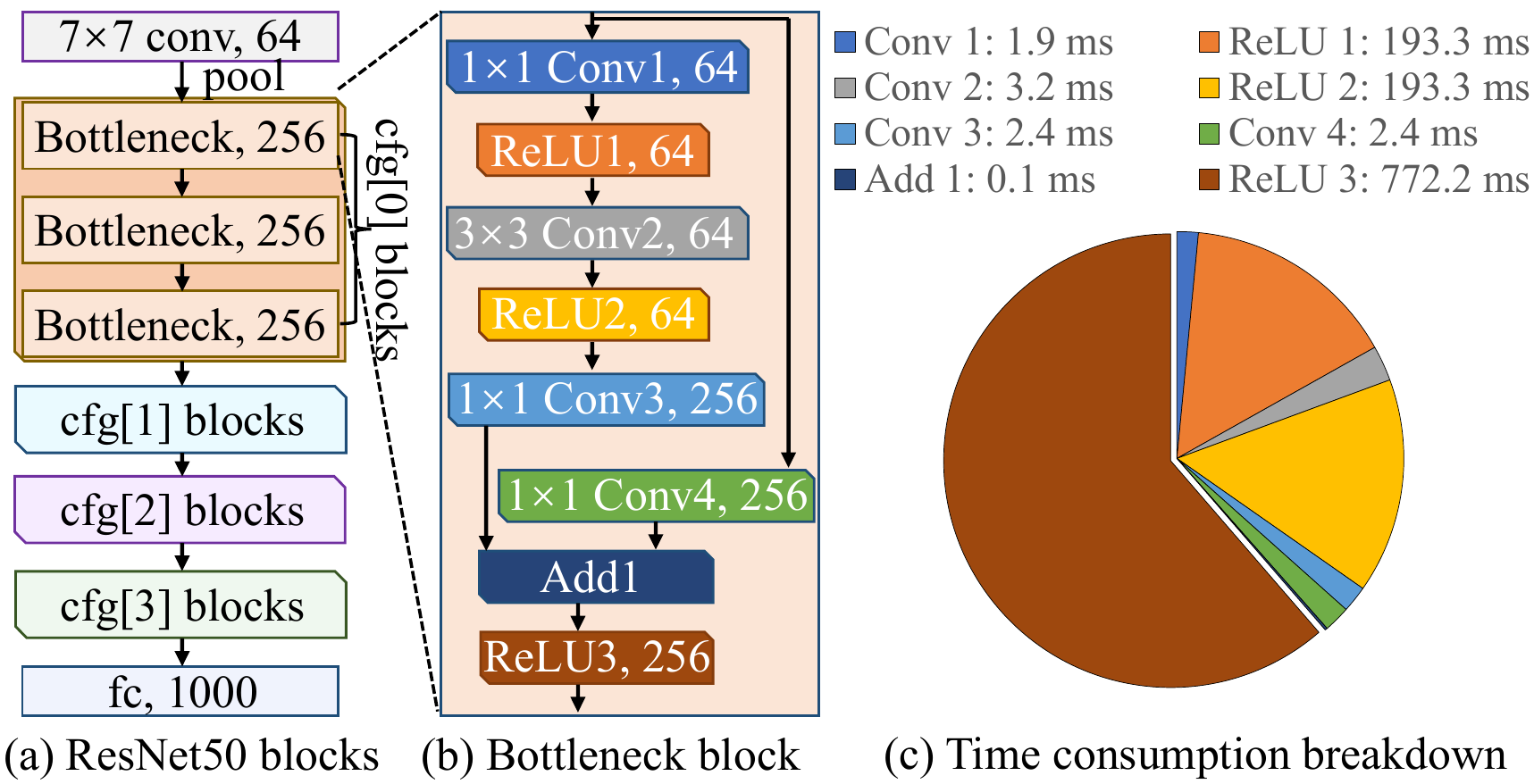}
  \captionof{figure}{Lantecy of operators under 2PC PI setup. Network banwidth: 1 GB/s. Device: ZCU104. Dataset: ImageNet.}
    \label{fig:motivation}
\end{figure}

The primary challenge in 2PC-based PI is the comparison protocol overhead~\cite{garay2007practical} for non-linear operators. As shown in Fig.~\ref{fig:motivation}, ReLU contributes over 99\% of latency in a ciphertext setting for deep neural network (DNN), despite negligible overhead in plaintext. Replacing ReLU with second-order polynomial activation could yield 50$\times$ speedup.


To achieve high performance, good scalability, and high energy efficiency for \textit{secure} deep learning systems, two orthogonal research directions have attracted enormous interest.  The first one is the nonlinear operations overhead reduction algorithms.
Existing works focus on \textit {ReLU cost optimization}, e.g., minimizing ReLU counts (DeepReduce~\cite{jha2021deepreduce}, CryptoNAS \cite{ghodsi2020cryptonas}) or replacing ReLUs with polynomials (CryptoNets~\cite{gilad2016cryptonets}, Delphi~\cite{mishra2020delphi}, SAFENet~\cite{lou2021safenet}), and \textit{extremely low-bit} weights and activations (e.g., Binary Neural Network (BNN) \cite{aggarwal2020soteria}). 
However, these works neglect the accuracy impact. They often sacrifice the model 
comprehension capability, resulting in several accuracy losses on large networks and datasets such as ImageNet, hence are not scalable.
The second trend is  {\em hardware acceleration for PI} to speed up the MPC-based DNN through GPUs~\cite{knott2021crypten, tan2021cryptgpu}. 
Since no hardware characteristic is captured during DNN design, this top-down ({"algorithm $\rightarrow$ hardware"}) approach can not effectively perform design space exploration, resulting in sub-optimal solutions. 

We focus on three observations: 1) preserving \textbf{prediction accuracy} for substantial benefits; 2) scalable \textbf{cryptographic overhead reduction} for various network sizes; 3) cohesive \textbf{algorithm/hardware optimizations} using closed loop "algorithm $\leftrightarrow$ hardware" with design space exploration capturing hardware characteristics.



We introduce the \textbf{Polynomial Architecture Search (PASNet)} framework, which jointly optimizes DNN model structure and hardware architecture for high-performance MPC-based PI. Considering cryptographic DNN operators, data exchange, and factors like encoding format, network speed, hardware architecture, and DNN structure, PASNet effectively enhances the performance of MPC-based PI.

Our key design principle is to {\em enforce} exactly what is assumed in the DNN design---training a DNN that is both hardware efficient and secure while maintaining high accuracy.

To evaluate the effectiveness of our framework, we use FPGA accelerator design as a demonstration due to its predictable performance, low latency, and high energy efficiency for MLaaS applications (e.g., Microsoft Azure~\cite{barnes2015azure}). 
We summarize our contributions as follows:





\leftmargini=4mm
\begin{enumerate}
\item  
We propose a trainable \textit{straight through polynomial activation initialization}  method for cryptographic hardware-friendly trainable polynomial activation function to replace the expensive ReLU operators. 
\item   Cryptographic hardware scheduler and the corresponding performance model are developed for the FPGA platform. The latency loop-up table is constructed. 
\item   We propose a differentiable cryptographic hardware-aware NAS framework to selectively choose the proper polynomial or non-polynomial activation based on given constraint and latency of cryptographic operators. 
\end{enumerate}

 
\section{\textbf{Basic of Cryptographic Operators}}
\label{sec:Cry_Building_Block}

\subsection{\textbf{Secret Sharing}}


\noindent\textbf{2PC setup.} 
We consider a similar scheme involving two semi-honest in a MLaaS applications~\cite{demmler2015aby}, where two servers receive the confidential inputs from each other and invoke a two party computing protocol for secure evaluation. 

\begin{figure}[t]
    \centering
      \includegraphics[width=1\linewidth]{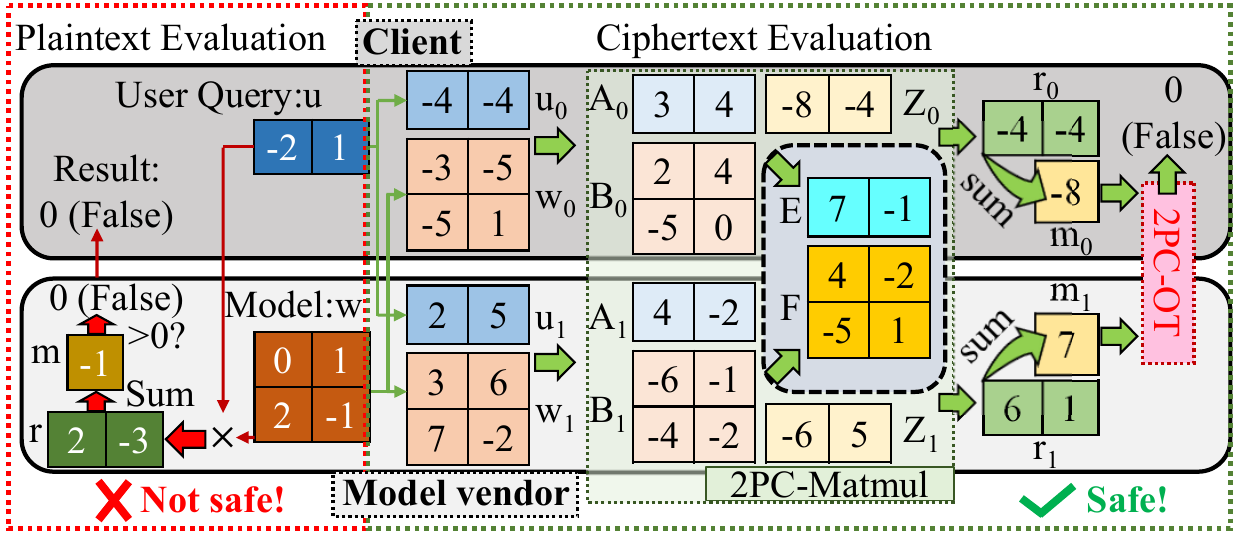}
  \captionof{figure}{A example of 4 bit plaintext vs. ciphertext evaluation. 
  }
    \label{fig:ss_example}
\end{figure}

\noindent\textbf{Additive Secret Sharing.} 
In this work, we evaluate 2PC secret sharing. 
As a symbolic representation, for a secret value $x\in \mathbb{Z}_m$, $\share{x}\gets(x_{S_0}, x_{S_1})$ denotes the two shares, where $x_{S_i}, i\in \{0,1\}$ belong to server $S_i$. Other notations are as below:
\begin{itemize}
    \item {\it Share Generation} $\mathbb{\textrm{shr}} (x)$: A random value $r$ in $\mathbb{Z}_{m}$ is sampled, and shares are generated as $\share{x}\gets (r, x-r)$.
    \item {\it Share Recovering}  $\mathbb{\textrm{rec}} ({\share{x}})$: Given shares $\share{x}\gets (x_{S_0}, x_{S_1})$, it computes $x\gets x_{S_0} + x_{S_1}$ to recover $x$.
\end{itemize}
An example of plaintext vs. secret shared based ciphertext evaluation is given in Fig.~\ref{fig:ss_example}, where ring size is 4 and $\mathbb{Z}_m =\{-8, -7, ... 7\}$. The integer overflow mechanism naturally ensures the correctness of ciphertext evaluation. Evaluation in the example involves secure multiplication, addition and comparison, and details are given in following sections. 




\subsection{\textbf{Polynomial Operators Over Secret-Shared Data}}\label{sec:poly}
\noindent\textbf{Scaling and Addition.} We denote secret shared matrices as $\share{X}$ and $\share{Y}$. The encrypted evaluation is given in Eq.~\ref{eq:mat_ad_ss}.

\begin{equation}\label{eq:mat_ad_ss}
\share{aX+Y}\gets(aX_{S_0}+Y_{S_0}, aX_{S_1}+Y_{S_1})
\end{equation}

\noindent\textbf{Multiplication.}
We consider the matrix multiplicative operations $\share{R}\gets \share{X} \otimes \share{Y}$ in the secret-sharing pattern.
where $\otimes$ is a general multiplication, such as Hadamard product, matrix multiplication, and convolution. 
We use oblivious transfer (OT)~\cite{kilian1988founding} based approach. 
To make the multiplicative computation secure, an extra Beaver triples~\cite{beaver1991efficient} should be generated as $\share{Z}=\share{A}\otimes\share{B}$, where $A$ and $B$ are randomly initialized. 
Specifically, their secret shares are denoted as $\share{Z}=(Z_{S_0}, Z_{S_1})$, $\share{A}=(A_{S_0}, A_{S_1})$, and $\share{B}=(B_{S_0}, B_{S_1})$. 
Later, two matrices are derived from given shares: $E_{S_i} = X_{S_i} - A_{S_i}$ and $F_{S_i} = Y_{S_i} - B_{S_i}$, in each party end separately. The intermediate shares are jointly recovered as $E\gets \mathbb{\textrm{rec}} {(\share{E})}$ and $F\gets \mathbb{\textrm{rec}} {(\share{F})}$. Finally, each party, i.e, server $S_i$, will calculate the secret-shared $R_{S_i}$ locally: 

\begin{equation}\label{eq:mat_mul_ss}
R_{S_i} = -i\cdot E \otimes F + X_{S_i} \otimes F  + E \otimes Y_{S_i} + Z_{S_i}
\end{equation}




\noindent\textbf{Square.}
For the element-wise square operator shown $\share{R}\gets \share{X} \otimes \share{X}$, we need to generate a Beaver pair $\share{Z}$ and $\share{A}$ where $\share{Z}=\share{A}\otimes\share{A}$, and $\share{A}$ is randomly initialized. 
Then parties evaluate $\share{E}=\share{X} - \share{A}$ and jointly recover $E\gets \mathbb{\textrm{rec}} {(\share{E})}$. The result $R$ can be obtained through Eq.~\ref{eq:sq_eval}. 

\begin{equation}\label{eq:sq_eval}
R_{S_i} = Z_{S_i} + 2 E \otimes A_{S_i}  + E \otimes E
\end{equation}

\subsection{\textbf{Non-Polynomial Operator Modules}}
Non-polynomial operators such as ReLU and MaxPool are evaluated using secure comparison protocol. \\
\noindent\textbf{Secure 2PC Comparison.}
The 2PC comparison, a.k.a. millionaires protocol, is committed to determine whose value held by two parties is larger, without disclosing the exact value to each other. 
We adopt work~\cite{garay2007practical} for 2PC comparison. 
Detailed modeling is given in Section~\ref{sec:hw-latency}.

\begin{figure*}[ht]
    \centering
      \includegraphics[width=.98\linewidth]{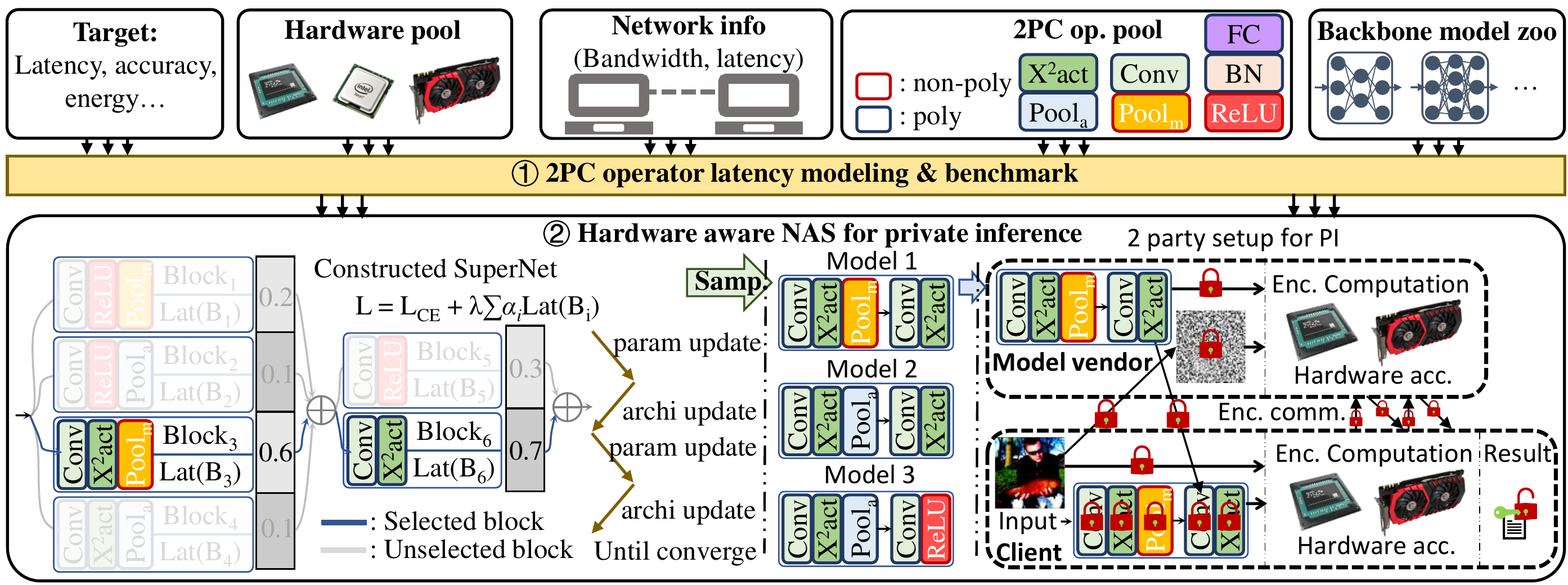}
  \captionof{figure}{Overview of PASNet framework for 2PC DNN based private inference setup. 
  }
    \label{fig:NAS}
\end{figure*}

\section{\textbf{The PASNet Framework}}




The framework (Fig.~\ref{fig:NAS}) takes inputs like optimization target, hardware pool, network information, and 2PC operator candidates for cryptographic operator modeling, benchmarking, and automated design space optimization in PI using hardware-aware NAS. This section presents a new cryptographic-friendly activation function, its initialization method, DNN operator modeling under 2PC, and a hardware-aware NAS framework for optimizing DNN accuracy and latency. While evaluated on FPGA accelerators, the method can be easily adapted to other platforms like mobile and cloud.

\subsection{\textbf{Trainable $X^{2}act$ Non-linear Function.}}

We use a hardware friendly trainable second order polynomial activation function as an non-linear function candidate, 
shown in Eq.~\ref{eq:x2act}, where $w_1$, $w_2$ and $b$ are all trainable parameters. 
We propose \textbf{\textit{straight through polynomial activation initialization} (STPAI)} method to set the $w_1$ and $b$ to be small enough and $w_2$ to be near to 1 in Eq.~\ref{eq:x2act} for initialization. 

\begin{equation}\label{eq:x2act}
\delta(x) = \frac{c}{\sqrt[]{N_x}} w_1 x^2 + w_2 x + b
\end{equation}





\noindent\textbf{Convergence.} Layer-wise second-order polynomial activation functions preserve the convexity of single-layer neural network \cite{sivaprasad2021curious}. Higher order polynomial activation function or channel-wise fine-grained polynomial replacement proposed in SAFENet \cite{lou2021safenet} may destroy the neural network's convexity and lead to a deteriorated performance. 

\noindent\textbf{Learning rate.} The gradient of $w_1$ must be balanced to match the update speed of other model weights. As such, we add a new scaling $\frac{c}{\sqrt[]{N_x}}$ prior to $w_1$ parameter. In the function, c is a constant, $N_x$ is the number of elements in feature map. 

\subsection{\textbf{Search Space of Hardware-aware NAS.}}\label{sec:searchspace}

We focus on convolutional neural networks (CNNs) in our study.
CNNs are mostly composed of Conv-Act-Pool and Conv-Act blocks. In work, we use the regular backbone model as a search baseline, such as the VGG family, mobilenetV3, and ResNet family. Each layer of supernet is composed of the layer structure obtained from baseline and its possible combination with $X^{2}act$ and $Pool_a$ replacement. A toy example is shown in Fig.~\ref{fig:NAS}, where a two-layer supernet is constructed, and the first layer is Conv-Act-Pool, and the second layer is Conv-Act. The first layer has four combinations which are Conv-ReLU-Pool\textsubscript{m}, Conv-ReLU-Pool\textsubscript{a}, Conv-$X^{2}act$-Pool\textsubscript{m}, and Conv-$X^{2}act$-Pool\textsubscript{a}. The second layer has two combinations: Conv-ReLU and Conv-$X^{2}act$. The Conv block's parameters can be either shared among candidates or separately trained during the search. 



\subsection{\textbf{Operator Modeling and Latency Analysis}}\label{sec:hw-latency}


This section will analyze five different operators: 2PC-ReLU, 2PC-$X^2act$, 2PC-MaxPool, 2PC-AvgPool, and 2PC-Conv. Therefore, they require $(1, n)$-OT (noted as \textbf{OT flow} block to implement  2PC comparison flows. Batch normalization can be fused into the convolution layer and it's not listed. 

\begin{figure}[t]
    \centering
      \includegraphics[width=1\linewidth]{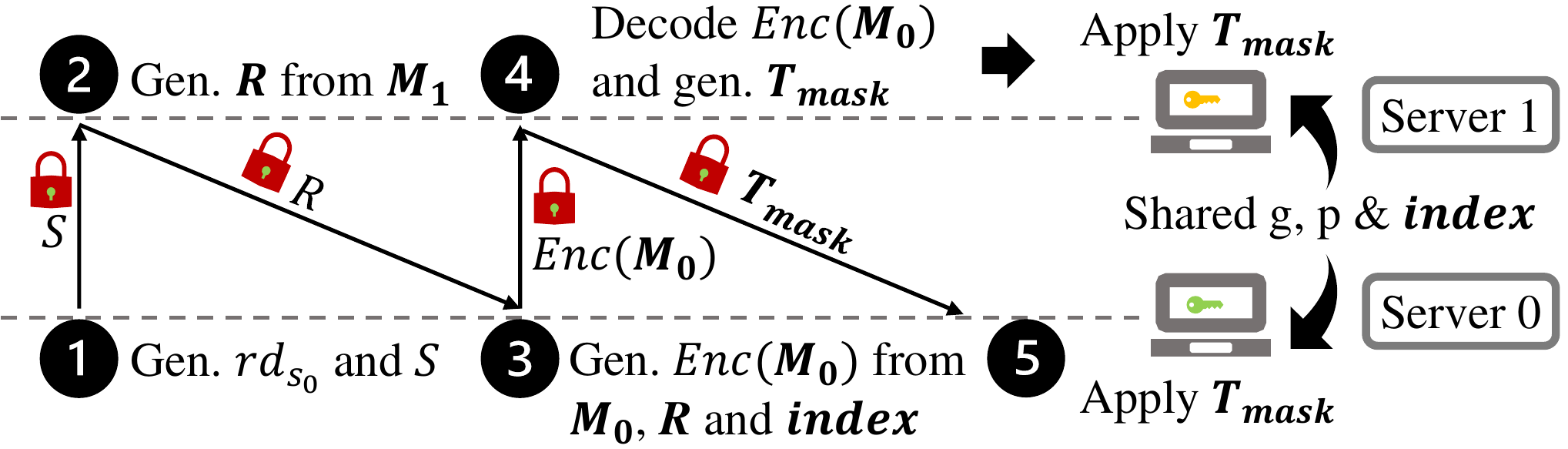}
  \captionof{figure}{Processing Steps of \textbf{2PC-OT} flow.}
    \label{fig:ot}
\end{figure}

\subsubsection{2PC-OT Processing Flow} While OT-based comparison protocol has been discussed in~\cite{kilian1988founding}, we hereby provide other communication detail as shown in Fig.~\ref{fig:ot}. 
Assume both servers have a shared prime number $m$, one generator ($g$) selected from the finite space $\mathbb{Z}_m$, and an \textbf{index} list with $L$ length. As we adopt 2-bit part, the length of \textbf{index} list is $L=4$. 

\noindent\textbf{\textcircled{1}} \textbf{Server 0} ($S_0$) generates a random integer $rd_{s_0}$, and compute mask number $S$ with $S = g^{rd_{S_0}}\ mod\ m$, then shares $S$ with the Server 1 ($S_1$). 
We only need to consider communication ($COMM_1$) latency as $COMM_1 = T_{bc} + \frac{32}{Rt_{bw}}$, since computation ($CMP_1$) latency is trivial.

\noindent\textbf{\textcircled{2}} \textbf{Server 1} ($S_1$) received $S$, and generates $\bm{R}$ list based on $S_1$'s 32-bit dataset $\bm{M_1}$, and then send them to $S_0$. Each element of $\bm{M_1}$ is split into $U = 16$ parts, thus each part is with 2 bits. 
Assuming the input feature is square with size $FI$ and $IC$ denotes the input channel, and we denote the computational parallelism as $PP$. The $CMP_2$ is modeled as Eq.~\ref{eq:CMP_2}, and $COMM_2$ is modeled as Eq.~\ref{eq:COMM_2}. 

\begin{equation}\label{eq:CMP_2}
    CMP_2 = \frac{32 \times 17 \times FI^2 \times IC}{PP \times freq}
    \vspace{-3pt}
\end{equation}   
\begin{equation}\label{eq:COMM_2}
    COMM_2 = T_{bc} + \frac{32 \times 16 \times FI^2 \times IC}{Rt_{bw}}
    \vspace{-3pt}
\end{equation}

\noindent\textbf{\textcircled{3}} \textbf{Server 0} ($S_0$) received $\bm{R}$, it will first generate the encryption $\bm{key_0}(y,u) = \bm{R}(y,u)\oplus (S^{b2d(\bm{M_1}(y, u)) + 1}\ mod\ m)^{rd_{S_0}}\ mod\ m$. The $S_0$ also generates is comparison matrix for it's $\bm{M_0}$ with 32-bit datatype and $U = 16$ parts, thus the matrix size for each value ($x$) is $4 \times 16$. The encrypted $Enc(\bm{M_0}(x,u))=\bm{M_0}(x,u)\oplus \bm{key_0}(y,u)$ will be sent to $S_1$. The $COMM_3$ of this step is shown in Eq.~\ref{eq:COMM_3}, and $CMP_3$ can be estimated as Eq.~\ref{eq:CMP_3}. 

\begin{equation}\label{eq:CMP_3}
    CMP_3 = \frac{32 \times (17 + (4 \times 16)) \times FI^2 \times IC}{PP \times freq}
    \vspace{-3pt}
\end{equation}    
\begin{equation}\label{eq:COMM_3}
    COMM_3 = T_{bc} + \frac{32 \times 4 \times 16 \times FI^2 \times IC}{Rt_{bw}}
\end{equation}

\noindent\textbf{\textcircled{4}} \textbf{Server 1} ($S_1$) decodes the interested encrypted massage by $\bm{key_1} = S^{rd_{S_0}}\ mod\ m$ in the final step. 
The $CMP_4$ and $COMM_4$ are calculated as following:

\begin{equation}\label{eq:CMP_4}
    CMP_4 = \frac{((32 \times 4 \times 16) + 1) \times FI^2 \times IC}{PP \times freq}
\end{equation}    
\begin{equation}\label{eq:COMM_4}
    COMM_4 = T_{bc} + \frac{FI^2 \times IC}{Rt_{bw}}
\end{equation}


\subsubsection{2PC-ReLU Operator} 2PC-ReLU requires 2PC-OT flow. 
2PC-ReLU latency ($Lat_{2PC-ReLu}$) model is given in Eq.~\ref{eq:LtReLU}.

\begin{equation}\label{eq:LtReLU}
    Lat_{2PC-ReLu} = \sum_{i = 2}^{4}CMP_i + \sum_{j = 1}^{4}COMM_j
\end{equation}

\subsubsection{2PC-MaxPool Operator} Original MaxPool function is shown in Eq. \ref{eq:org_maxpool}. The 2PC-MaxPool uses OT flow comparison, and the latency model is shown in Eq.~\ref{eq:LtMax}.



\begin{equation}\label{eq:org_maxpool}
    out = \max_{\substack{
k_h \in [0, K_h -1] \\
k_w \in [0, K_w -1]\\
}} in(n, c, hS_h + k_h, wS_w + k_w )
\end{equation}

\begin{equation}\label{eq:LtMax}
    Lat_{2PC-MaxPool} = \sum_{i = 2}^{4}CMP_i + \sum_{j = 1}^{4}COMM_j + 3T_{bc}
\end{equation}

\subsubsection{2PC-$X^2$act Operator} The original $X^2act$ has been shown in Eq.~\ref{eq:x2act}. 
The $X^2act$ needs a ciphertext square operation and 2 ciphertext-plaintext multiplication operations. The basic protocol is demonstrated in Sec.~\ref{sec:poly}. 
The latency of computation and communication can be modeled as: $CMP_{x^2} = \frac{2 \times FI^2 \times IC}{PP \times freq} $ and $ COMM_{x^2} = T_{bc} + \frac{32 \times FI^2 \times IC}{Rt_{bw}}$. The latency model of 2PC-$X^2$act ($Lat_{2PC-X^2act}$) is shown in Eq.~\ref{eq:LtX2}.

\begin{equation}\label{eq:LtX2}
    Lat_{2PC-X^2act} = CMP_{x^2} + 2 \times COMM_{x^2}
\end{equation}

\subsubsection{2PC-AvgPool Operator} The 2PC-AvgPool operator only involves addition and scaling, the latency is 

\begin{equation}
    Lat_{2PC-AvgPool} = \frac{2 \times FI^2 \times IC}{PP \times freq}
\end{equation}

\subsubsection{2PC-Conv Operator} The 2PC-Conv operator involves multiplication between ciphertext, and the basic computation and communication pattern are given Sec.~\ref{sec:poly}. The computation part follows tiled architecture implementation \cite{zhang2015optimizing}. 
Assuming we can meet the computation roof by adjusting tiling parameters, the latency of the 2PC-Conv computation part can be estimated as $CMP_{Conv} = \frac{3 \times K \times K \times FO^2 \times IC \times OC}{PP \times freq}$, where $K$ is the convolution kernel size. The communication latency is modeled as $ COMM_{Conv} = T_{bc} + \frac{32 \times FI^2 \times IC}{Rt_{bw}}$. Thus, the latency of 2PC-Conv is given in Eq.~\ref{eq:Ltconv}. 

\begin{equation}\label{eq:Ltconv}
    Lat_{2PC-Conv} = CMP_{Conv} + 2 \times COMM_{Conv}
\end{equation}

\subsection{\textbf{Differentiable Harware Aware NAS Algorithm}}

\label{sec:archi_search}

\begin{algorithm}[htb] 
\small
\caption{Differentiable Polynomial Architecture Search. 
} 
\label{alg:framework} 
\begin{algorithmic}[1] 
\REQUIRE 
$M_b$: backbone model; $D$: a specific dataset\\
~~~~ $Lat(OP)$: latency loop up table; $H$: hardware resource
\ENSURE 
Searched polynomial model $M_p$
\WHILE{not converged}
 \STATE Sample minibatch $x_{trn}$ and $x_{val}$ from trn. and val. dataset
 \STATE // Update architecture parameter $\alpha$:
 \STATE Forward path to compute $\zeta_{trn}(\omega, \alpha)$ based on $x_{trn}$
 \STATE Backward path to compute $\delta \omega = \frac{\partial \zeta_{trn}(\omega, \alpha)}{\partial \omega}$
 \STATE Virtual step to compute $ \omega' = \omega - \xi\delta\omega$
 \STATE Forward path to compute $\zeta_{val}(\omega', \alpha)$ based on $x_{val}$
 \STATE Backward path to compute $\delta \alpha' = \frac{\partial \zeta_{val}(\omega', \alpha)}{\partial \alpha}$
 \STATE Backward path to compute $\delta \omega' = \frac{\partial \zeta_{val}(\omega', \alpha)}{\partial \omega'}$
 \STATE Virtual steps to compute $ \omega^{\pm} = \omega \pm \varepsilon\delta\omega'$ 
 \STATE Two forward path to compute $\zeta_{trn}(\omega^{\pm}, \alpha)$ 
 \STATE Two backward path to compute $\delta \alpha^{\pm} =  \frac{\partial \zeta_{trn}(\omega^{\pm}, \alpha)}{\partial \alpha}$ 
 \STATE Compute hessian $\delta \alpha'' = \frac{\delta \alpha^{+} - \delta \alpha^{-}}{2\varepsilon}$
 \STATE Compute final architecture parameter gradient $\delta\alpha = \delta \alpha'- \xi\delta \alpha'' $
 \STATE Update architecture parameter using $\delta\alpha$ with Adam optimizer
 \STATE // Update weight parameter $\omega$:
 \STATE Forward path to compute $\zeta_{trn}(\omega, \alpha)$ based on $x_{trn}$
 \STATE Backward path to compute $\delta \omega = \frac{\partial \zeta_{trn}(\omega, \alpha)}{\partial \omega}$
 \STATE Update architecture parameter using $\delta\omega$ with SGD optimizer
\ENDWHILE
\\ Obtain architecture by $OP_{l}(x) = OP_{l, k^*}(x), \: s.t. \: k^* = \mathbb{\textrm{argmax}}_{k} \: \theta_{l, k}$
\end{algorithmic}
\end{algorithm}

\begin{figure*}[ht!]
    \centering
    \centering
\begin{multicols}{2}
\subfloat [\label{fig:hardware_throughput}Searched model accuracy comparison]   {\includegraphics[width=0.92\columnwidth]{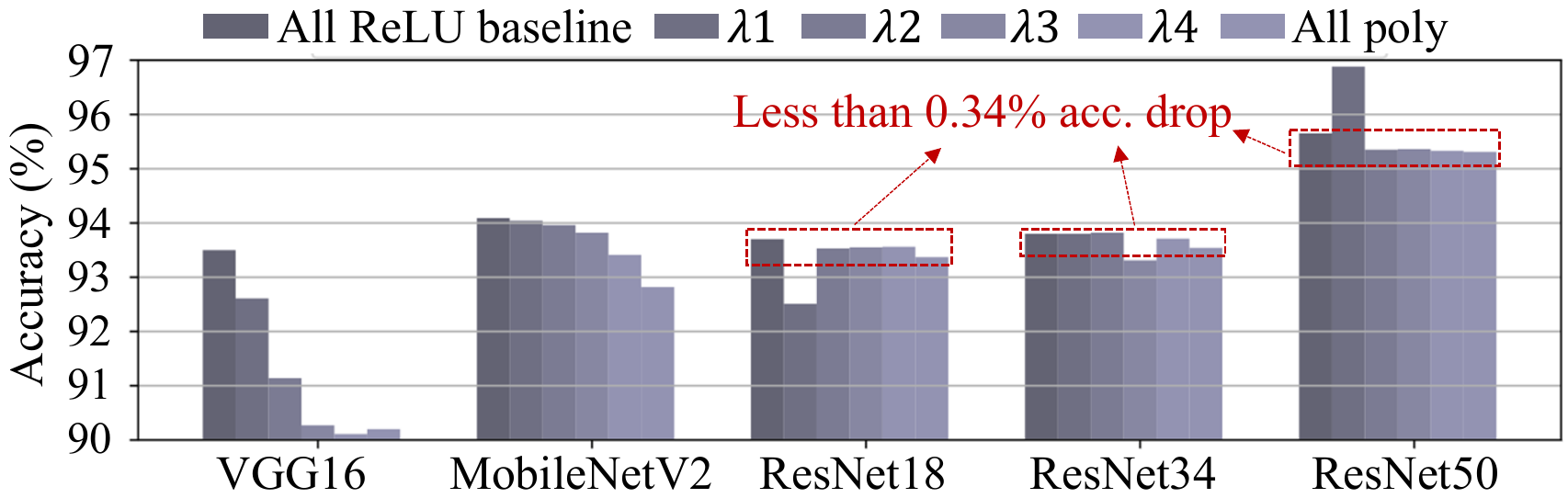}\par }
\subfloat  [\label{fig:attention_throughput}Searched model private inference latency comparison]  {\hspace{.3in}\includegraphics[width=0.9\columnwidth]{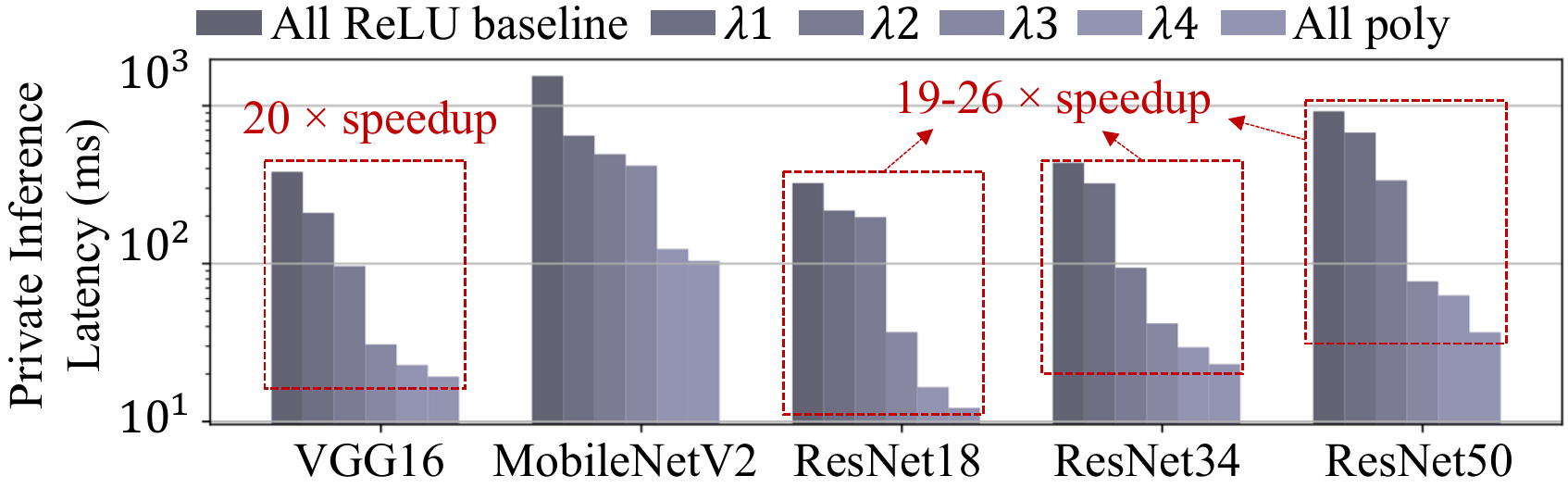}\par }
\end{multicols}

    \caption{PASNet framework evaluation on CIFAR-10 dataset under 2PC PI setup. Network banwidth: 1 GB/s. Device: ZCU104. }
    \label{fig:pasnet_cifar10}
\end{figure*}

Early work \cite{tan2019efficientnet} focus on using RL for NAS. The RL based method effectively explores the search space but still requires a significant amount of search overhead such as GPU hours and energy. 
Hardware-aware NAS have also been investigated \cite{wu2019fbnet}.
In this work, we incorporate latency constraint into the target loss function of the DARTS framework~\cite{liu2018darts}, and develop a differentiable cryptographic hardware-aware micro-architecture search framework. We firstly determine a supernet model for NAS, and introduces gated operators $OP_{l}(x)$ which parametrizes the candidate operators $OP_{l, j}(x)$ selection with a trainable weight $\alpha_{l,k}$ (Eq.~\ref{eq:softmax_para}). For example, a gated pooling operator consists of MaxPool and AvgPool operators and 2 trainable parameters for pooling selection. The latency of the operators could be determined based on Sec.~\ref{sec:hw-latency}. A parameterized latency constraint is given as $Lat(\alpha) = \sum_{l = 1}^{n}\sum_{j = 1}^{m}\theta_{l, j}Lat(OP_{l, j})$, 
where the latency of gated operators are weighted by $\theta_{l, j}$. 
We incorporate the latency constraint into the loss function as $\zeta(\omega, \alpha) = \zeta_{CE}(\omega, \alpha) + \lambda Lat(\alpha)$, and penalize the latency $Lat(\alpha)$ by $\lambda$. 




\begin{equation}\label{eq:softmax_para}
\theta_{l, j} = \frac{\exp(\alpha_{l,j})}{\sum_{k = 1}^{m}\exp(\alpha_{l,k})}, \: OP_{l}(x) = \sum_{k = 1}^{m} \theta_{l, k}OP_{l, k}(x)
\end{equation}


The optimization objective of our design is shown in Eq.~\ref{eq:optimization}, we aim to minimize the validation loss $\zeta_{val}(\omega^*, \alpha)$ with regard to architecture parameter $\alpha$. The optimal weight $\omega^*$ is obtained through minimize the training loss. The second order approximation of the optimal weight is given as $\omega^* \approx \omega' = \omega - \xi \: \delta \zeta_{trn}(\omega, \alpha)/\delta\omega$,
the approximation is based on current weight parameter and its' gradient. The virtual learning rate $\xi$ can be set equal to that of weight optimizer. 


\begin{equation}\label{eq:optimization}
\mathbb{\textrm{argmin}}_{\alpha} \: \zeta_{val}(\omega^*, \alpha), \: s.t. \: \omega^* = \mathbb{\textrm{argmin}}_{\omega} \: \zeta_{trn}(\omega, \alpha)
\end{equation}

Eq.~\ref{eq:opt_2} gives the approximate $\alpha$ gradient using chain rule, the second term of $\alpha$ gradient can be further approximated using small turbulence $\varepsilon$, where weights are $\omega^{\pm} = \omega \pm \varepsilon \: \delta\zeta_{val}(\omega', \alpha)/\delta\omega'$ 
and Eq.~\ref{eq:hession} is used for final $\alpha$ gradient. 

\begin{equation}\label{eq:opt_2}
\delta\zeta_{val}(\omega', \alpha)/\delta\alpha - \xi  \: \delta\zeta_{val}(\omega', \alpha)/\delta\omega' \: \delta\delta\zeta_{trn}(\omega, \alpha)/\delta\omega\delta\alpha
\end{equation}
\begin{equation}\label{eq:hession}
\frac{\delta\delta\zeta_{trn}(\omega, \alpha)}{\delta\omega\delta\alpha}
 =  \delta(\zeta_{trn}(\omega^+, \alpha) - \zeta_{trn}(\omega^-, \alpha))/(2\varepsilon\delta\alpha)
\end{equation}

With the help of analytical modeling of optimization objective, we are able to derive the differentiable polynomial architecture search framework in Algo.~\ref{alg:framework}. The input of search framework includes backbone model $M_b$, dataset $D$, latency loop up table $Lat(OP)$, and hardware resource $H$. The algorithm returns a searched polynomial model $M_p$. The algorithm iteratively trains the architecture parameter $\alpha$ and weight $\omega$ parameter till the convergence. Each $\alpha$ update requires 4 forward paths and 5 backward paths according to Eq.~\ref{eq:optimization} to Eq.~\ref{eq:hession}, and each $\omega$ update needs 1 forward paths and 1 backward paths. After the convergence of training loop, the algorithm returns a deterministic model architecture by applying $OP_{l}(x) = OP_{l, k^*}(x), \: s.t. \: k^* = \mathbb{\textrm{argmax}}_{k} \: \alpha_{l, k}$. The returned architecture is then used for 2PC based PI evaluation. 

\section{\textbf{Evaluation}}

\noindent\textbf{Hardware setup.} Our platform uses two ZCU104 MPSoCs connected via a 1 GB/s LAN router. With a 128-bit load/store bus and 32-bit data, we process four data simultaneously at 200MHz. The fixed point ring size is set to 32 bits for PI.

\noindent\textbf{Datasets and Backbone Models. }
PASNet is evaluated on CIFAR-10 and ImageNet for image classification tasks. CIFAR-10~\cite{krizhevsky2017imagenet} has colored $32 \times 32$ images, with $10$ classes, $50,000$ training, and $10,000$ validation images. ImageNet~\cite{krizhevsky2017imagenet} has RGB $224 \times 224$ images, with $1000$ categories, $1.2$ million training, and $50,000$ validation images.



\noindent\textbf{Systems Setup. }
Polynomial architecture search experiments are conducted using Ubuntu 18.04, Nvidia Quadro RTX 6000 GPU, PyTorch v1.8.1, and Python 3.9.7. Pretrained weights for CIFAR-10 and ImageNet are from~\cite{huy_phan_2021_4431043} and Pytorch Hub~\cite{pytorch_hub}, respectively. Cryptographic DNN inference is performed on FPGA-based accelerators using two ZCU104 boards, connected via Ethernet LAN. The FPGA accelerators are optimized with coarse-grained and fine-grained pipeline structures, as discussed in Sec.~\ref{sec:hw-latency}.


\begin{table*}[]
\caption{PASNet evaluation \& cross-work comparison with CryptGPU \cite{tan2021cryptgpu} and CryptFLOW \cite{kumar2020cryptflow}. Batch size = 1}
\centering
\resizebox{0.9\linewidth}{!}{
\begin{tabular}{c|cccc|ccccc}
\hline
                                                             & \multicolumn{4}{c|}{CIFAR-10 dataset}                                                                                                    & \multicolumn{5}{c}{ImageNet dataset}                                                                                                                   \\ \hline
Model                                                        & \multicolumn{1}{c|}{Top 1 (\%)}       & \multicolumn{1}{c|}{Lat.  (ms)}        & \multicolumn{1}{c|}{Comm. (MB)}       & Effi. (1/(ms*kW)) & \multicolumn{1}{c|}{Top 1 (\%)} & \multicolumn{1}{c|}{Top 5 (\%)} & \multicolumn{1}{c|}{Lat. (s)} & \multicolumn{1}{c|}{Comm. (GB)} & Effi. (1/(s*kW)) \\ \hline
PASNet-A                                                     & \multicolumn{1}{c|}{93.37}            & \multicolumn{1}{c|}{12.2}              & \multicolumn{1}{c|}{2.86}           & 5.12             & \multicolumn{1}{c|}{70.54}      & \multicolumn{1}{c|}{89.59}      & \multicolumn{1}{c|}{0.063}      & \multicolumn{1}{c|}{0.035}     & 999             \\ \hline
PASNet-B                                                     & \multicolumn{1}{c|}{95.31}            & \multicolumn{1}{c|}{36.74}              & \multicolumn{1}{c|}{13.18}              & 1.70              & \multicolumn{1}{c|}{78.79}      & \multicolumn{1}{c|}{93.99}      & \multicolumn{1}{c|}{0.228}      & \multicolumn{1}{c|}{0.162}        & 274              \\ \hline
PASNet-C                                                     & \multicolumn{1}{c|}{95.33}            & \multicolumn{1}{c|}{62.91}              & \multicolumn{1}{c|}{30.03}              & 0.99              & \multicolumn{1}{c|}{79.25}      & \multicolumn{1}{c|}{94.38}      & \multicolumn{1}{c|}{0.539}      & \multicolumn{1}{c|}{0.368}        & 115              \\ \hline
PASNet-D                                                     & \multicolumn{1}{c|}{92.82}            & \multicolumn{1}{c|}{104.09}             & \multicolumn{1}{c|}{25.01}           & 0.60              & \multicolumn{1}{c|}{71.36}      & \multicolumn{1}{c|}{90.15}      & \multicolumn{1}{c|}{0.184}     & \multicolumn{1}{c|}{0.103}      & 339              \\ \hline
\begin{tabular}[c]{@{}c@{}}CryptGPU\\ ResNet50\end{tabular}  & \multicolumn{1}{c|}{\textbackslash{}} & \multicolumn{1}{c|}{\textbackslash{}} & \multicolumn{1}{c|}{\textbackslash{}} & \textbackslash{} & \multicolumn{1}{c|}{78}         & \multicolumn{1}{c|}{92}         & \multicolumn{1}{c|}{9.31}     & \multicolumn{1}{c|}{3.08}       & 0.15             \\ \hline
\begin{tabular}[c]{@{}c@{}}CryptFLOW\\ ResNet50\end{tabular} & \multicolumn{1}{c|}{\textbackslash{}} & \multicolumn{1}{c|}{\textbackslash{}} & \multicolumn{1}{c|}{\textbackslash{}} & \textbackslash{} & \multicolumn{1}{c|}{76.45}      & \multicolumn{1}{c|}{93.23}      & \multicolumn{1}{c|}{25.9}     & \multicolumn{1}{c|}{6.9}        & 0.096            \\ \hline
\end{tabular}
}
\label{tab:pasnet_eval}
\end{table*}

\subsection{\textbf{Hardware-aware NAS Evaluation}}
Our hardware-aware PASNet evaluation experiment (algorithm descripted in Sec.~\ref{sec:archi_search}) was conducted on CIFAR-10 training dataset. A new training \& validation dataset is randomly sampled from the CIFAR-10 training dataset with 50\%-50\% split ratio. The new training dataset is used to update the weight parameter of PASNet models, and the new validation dataset is used to update the architecture parameter. 

The hardware latency is modeled through section.~\ref{sec:hw-latency}, and the $\lambda$ for latency constraint in loss function is tuned to generate architectures with different latency-accuracy trade-off. 
Prior search starts, the major model parameters are randomly initialized and the polynomial activation function is initialized through \textbf{STPAI} method. 
We use VGG-16~\cite{simonyan2014very}, ResNet-18, ResNet-34, ResNet-50~\cite{he2016deep}, and MobileNetV2~\cite{sandler2018mobilenetv2} as backbone model structure to evaluate our PASNet framework. 


With the increase of latency penalty, the searched structure's accuracy decreases since the DNN structure has more polynomial operators. After the proper model structure is found during architecture search process, the transfer learning with \textbf{STPAI} is conducted to evaluate the finetuned model accuracy.

The finetuned model accuracy under 2PC setting with regard to $\lambda$ setting can be found in Fig.~\ref{fig:pasnet_cifar10}(a). The baseline model with all ReLU setting and all-polynomial operation based model are also included in the figure for comparison. Generally, a higher polynomial replacement ratio leads to a lower accuracy. The VGG-16 model is the most vulnerable model in the study, while the complete polynomial replacement leads to a 3.2\% accuracy degradation (baseline 93.5\%). On the other side, ResNet family are very robust to full polynomial replacement and there are only $0.26\%$ to $0.34\%$ accuracy drop for ResNet-18 (baseline 93.7\%), ResNet-34 (baseline 93.8\%) and ResNet-50 (baseline 95.6\%). MobileNetV2's is in between the performance of VGG and ResNet, in which a full polynomial replacement leads to $1.27\%$ degradation (baseline 94.09\%).

On the other hand, Fig.~\ref{fig:pasnet_cifar10}(b) presents the latency profiling result of searched models performance on CIFAR-10 dataset under 2PC setting. All polynomial replacement leads to 20 times speedup on VGG-16 (baseline 382 ms), 15 times speedup on MobileNetV2 (baseline 1543 ms), 26 times speedup, ResNet-18 (baseline 324 ms), 19 times speedup on ResNet-34 (baseline 435 ms), and 25 times on speedup ResNet-50 (baseline 922 ms). With most strict constraint $\lambda$, the searched model latency is lower.  

\subsection{\textbf{Cross-work ReLU Reduction Performance Comparison}}

\begin{figure}[htpb!]
\centering
\includegraphics[width=0.9\linewidth]{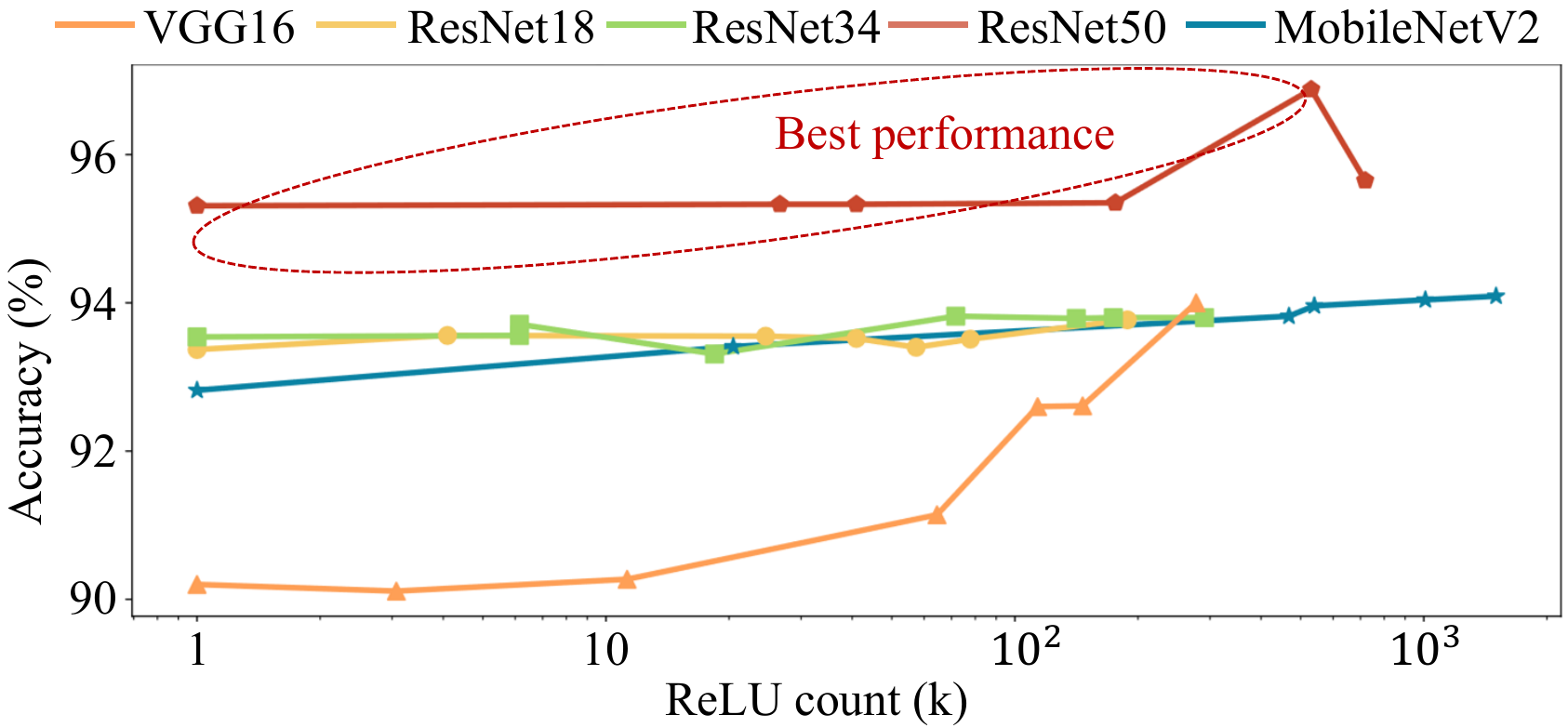}

\caption{Accuracy-ReLU count trade-off on CIFAR-10. }
\label{fig:nas_acc_pareto}

\end{figure}

\begin{figure} [htpb!]
\centering
\includegraphics[width = 0.9\linewidth]{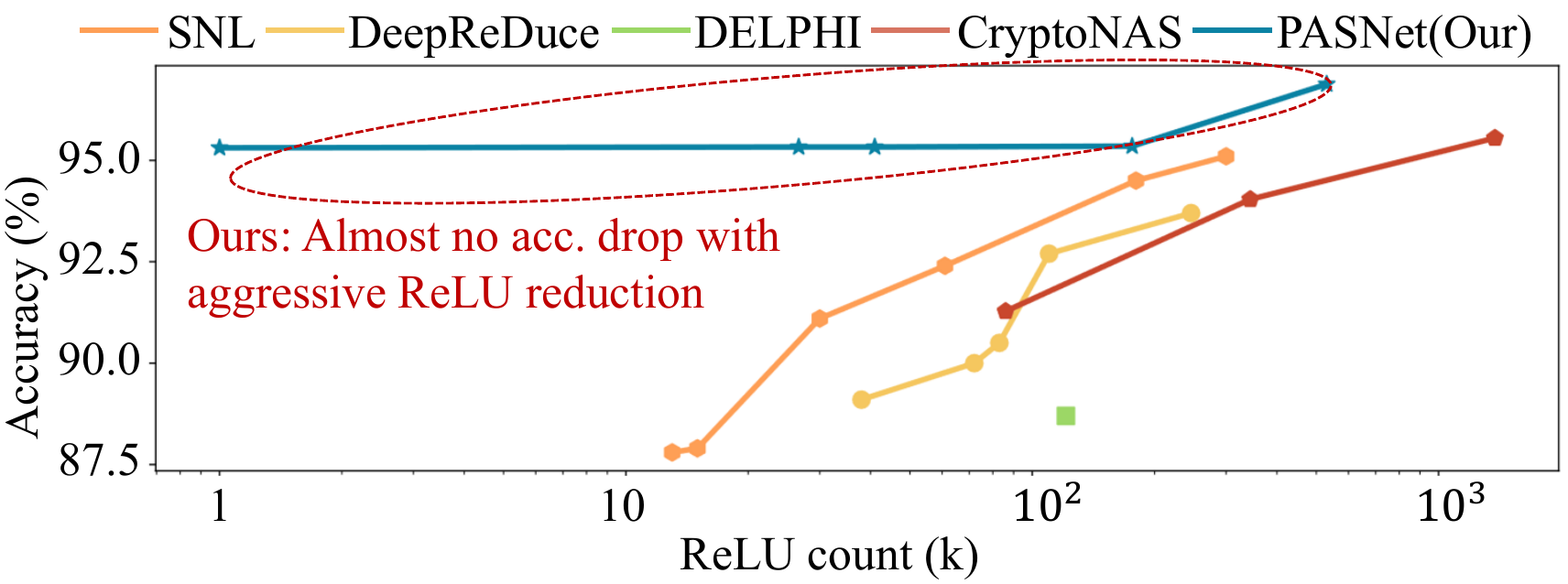}
\caption{ReLU reduction comparison on CIFAR-10. }
\label{fig:algo_cross_work}

\end{figure}

A futher accuracy-ReLU count analysis is conducted and compared with SOTA works with ReLU reduction: DeepReDuce \cite{jha2021deepreduce}, DELPHI \cite{mishra2020delphi}, CryptoNAS \cite{ghodsi2020cryptonas}, and SNI \cite{cho2022selective}. As shown in Fig.~\ref{fig:nas_acc_pareto}, we generate the pareto frontier with best accuracy-ReLU count trade-off from our architecture search result. We name the selected models as \textbf{PASNet}, and compare it with other works. The accuracy-ReLU count comparison is show in Fig.~\ref{fig:algo_cross_work}. Our work achieves a much better accuracy vs. ReLU comparison than existing works, especially at the situation with extremely few ReLU counts. 

\subsection{\textbf{Cross-work PI System Performance Comparison}}

We pick up 4 searched PASNet model variants for CIFAR-10 \& ImageNet dataset accuracy \& latency evaluation and name them as \textbf{PASNet-A}, \textbf{PASNet-B}, \textbf{PASNet-C}, \textbf{PASNet-D}. PASNet-A is a light-weighted model and shares the same backbone model as ResNet-18 but has only polynomial operators.
PASNet-B and PASNet-C are heavily-weighted models that share the same backbone model as ResNet-50. PASNet-B has only polynomial operators and PASNet-C has 4 2PC-ReLU operators. 
PASNet-D is a medium-weighted model derived from MobileNetV2 with all polynomial layers. 
Note that the baseline top-1 accuracy of ResNet-18 on CIFAR-10 and ImageNet are 93.7\% and 69.76\%, baseline top-1 accuracy of ResNet-50 on CIFAR-10 and ImageNet are 95.65\% and 78.8\%, and the baseline top-1 accuracy of MobileNetV2 on CIFAR-10 and ImageNet are 94.09\% and 71.88\%. 

The PASNet variants evaluation results and ImageNet cross-work comparison with SOTA CryptGPU~\cite{tan2021cryptgpu} and CryptFLOW~\cite{kumar2020cryptflow} implementation can be found in Tab.~\ref{tab:pasnet_eval}.  We observe a 0.78\% top-1 accuracy increase for our light-weighted PASNet-A compared to baseline ResNet-18 performance on ImageNet. Heavily-weighted models PASNet-B and PASNet-C achieve comparable (-0.01\%) or even higher accuracy (+0.45\%) than the ResNet-50 baseline. we achieve only a 0.13\% accuracy drop for our medium-weighted PASNet-D compared to baseline MobileNetV2 performance on ImageNet. Even with the ZCU 104 edge devices setting, we can achieve a much faster secure inference latency than the SOTA works implemented on the large-scale server system. Our light-weighted PASNet-A achieves 147 times latency reduction and 88 times communication volume reduction compared to CryptGPU~\cite{tan2021cryptgpu}. Our heavily-weighted model PASNet-B achieved 40 times latency reduction and 19 times communication volume reduction than CryptGPU~\cite{tan2021cryptgpu} while maintaining an even higher accuracy. Our highest accuracy model PASNet-C achieved 79.25\% top-1 accuracy on the ImageNet dataset with 17 times latency reduction and 8.3 times communication volume reduction than CryptGPU~\cite{tan2021cryptgpu}. Note that our system is built upon the ZCU104 edge platform, so our energy efficiency is much higher (more than 1000 times) than SOTA CryptGPU~\cite{tan2021cryptgpu} and CryptFLOW~\cite{kumar2020cryptflow} systems. 



\section{\textbf{Discussion}}


Existing MLaaS accelerations focused on plaintext inference acceleration~\cite{peng2022length, huang2021sparse, zhang2023accelerating, zhang2022toward, wu2020intermittent, bao2019efficient, peng2022towards, kan2021zero, luo2022codg, bao2020fast, li2022makes, peng2021accelerating, kan2022brain, qi2021accommodating, xiao2019autoprune, peng2021binary, wang2021lightweight, huang2021hmc, kan2022fbnetgen, zhang2022algorithm, qi2021accelerating, sheng2022larger, li2021generic, huang2022automatic}. Others target on plaintext training acceleration~\cite{wu2022decentralized, huang2023neurogenesis, wu2020enabling, xu2023neighborhood, wu2023synthetic, bao2022doubly, wu2021enabling, huang2022dynamic, wu2022distributed, bao2022accelerated, lei2023balance}, federated learning~\cite{wu2021federated1, wang2022variance, wu2021federated2} to protect the privacy of training data, and privacy protection of model vendor~\cite{wang2020against, wang2022analyzing}.

In this work, we propose PASNet to reduce high comparison protocol overhead in 2PC-based privacy-preserving DL, enabling low latency, high energy efficiency, and accurate 2PC-DL. We employ hardware-aware NAS with latency modeling. Experiments demonstrate PASNet-A and PASNet-B achieve 147x and 40x speedup over SOTA CryptGPU on ImageNet PI test, with 70.54\% and 78.79\% accuracy.

\section*{Acknowledgement}
This work  was  in  part  supported  by the NSF CNS-2247891, 2247892, 2247893, CNS-2153690, DGE-2043183, and the Heterogeneous Accelerated Compute Clusters (HACC) program at UIUC. Any opinions, findings and conclusions, or recommendations expressed in this material are those of the authors and do not necessarily reflect the views of the funding agencies.

\bibliographystyle{unsrt}
\bibliography{ref}

\end{document}